\documentclass{IEEE_lsens}
\usepackage{textcomp}
\usepackage[noadjust]{cite}
\ifCLASSINFOpdf
\else
\fi
\usepackage[T1]{fontenc} 
\usepackage{amsmath}
\usepackage[cmintegrals]{newtxmath}
\usepackage{graphicx}
\usepackage{units}
\usepackage{bm}
\usepackage{array}
\usepackage{url}

\ifCLASSINFOpdf
\else
\fi
\providecommand{\hypersetup}[1]{\relax}

\hypersetup{pdftitle={Analysis of attitude errors in GRACE range-rate residuals - a comparison between SCA1B and the reprocessed attitude fused product (SCA1B + ACC1B)},
pdfsubject={Typesetting},
pdfauthor={Sujata Goswami},
pdfkeywords={GRACE, residual analysis, IEEE\_lsens, IEEE Sensors Letters, LaTeX, Typesetting, TeX}}
\begin{document}

\markboth{Vol.~1, No.~3, July~2017}{0000000}

\IEEELSENSarticlesubject{Sensor Applications}

%
\title{Analysis of attitude errors in GRACE range-rate residuals - a comparison between SCA1B and the reprocessed attitude fused product (SCA1B + ACC1B)}
%
\author{\IEEEauthorblockN{Sujata~Goswami\IEEEauthorrefmark{*}}
\IEEEauthorblockA{\IEEEauthorrefmark{*}Institut F\"ur Erdmessung,
Schneiderberg 50, 30167, Leibniz University of Hannover, Germany}
\thanks{Corresponding author: S. Goswami (e-mail: sujata.goswami@aei.mpg.de).}
\thanks{Associate Editor: Alan Smithee.}%
\thanks{Digital Object Identifier 10.1109/LSENS.2017.0000000}}
%
%
%


\IEEEtitleabstractindextext{%
\begin{abstract}
  The precision of the attitude in the inter-satellite ranging missions like \textsc{grace} is one of the important requirement. It is required not only for the mission performance but also for the good quality of the gravity field models which are estimated from the inter-satellite ranging measurements. Here we present a comparative study of the analysis of two attitude datasets. One of them is the standard \textsc{sca1b} release 2 datasets provided by \textsc{jpl nasa} and another is the reprocessed attitude computed at \textsc{tu g}raz by combining the angular accelerations and the standard \textsc{sca1b} release 2 datasets. Further we also present the impact of the attitude datasets on the inter-satellite range measurements by analyzing their residuals. Our analysis reveals the significant improvement in the attitude due to the reprocessed product and reduced value of residuals computed from the reprocessed attitude.

\end{abstract}

}

\maketitle
\section{Introduction}
\begin{figure}[htbp!]
  \centering
\includegraphics[width = 0.78\linewidth]{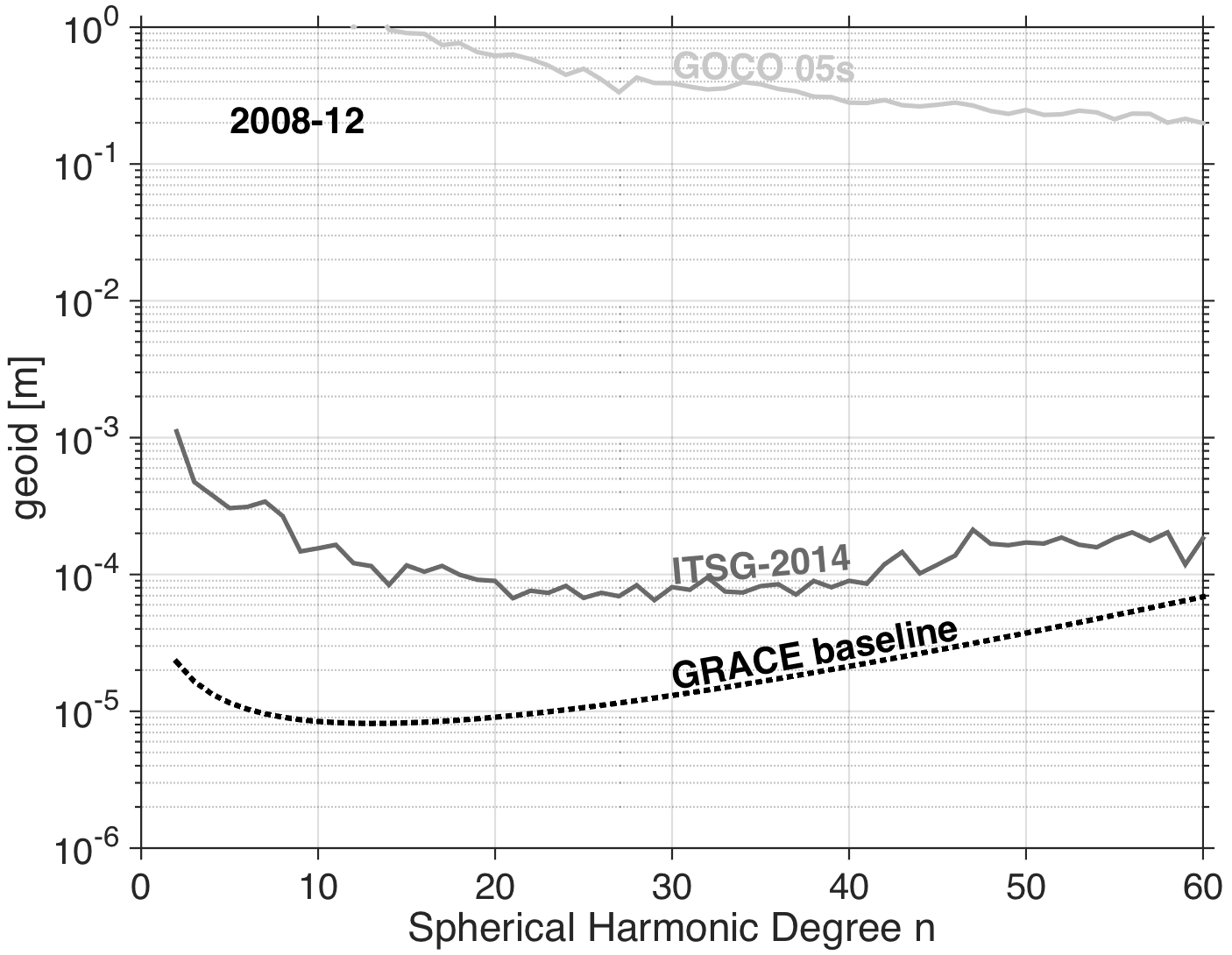}
\caption{Geoid degree amplitudes of the \textsc{itsg}-2014 solutions compared with the \textsc{grace} baseline. The differences are presented with respect to \textsc{goco}05s static field. The gravity field solution is for the month of \textsc{d}ecember 2008. \label{figBaseline}}
\end{figure}

The \textsc{grace} satellite mission has successfully provided the gravity field products for more than 15 years (2002--2017).
The gravity field solutions are computed from the \textsc{k}-band range-rate observations ($\dot\rho$) which measures the tiny mass changes in the \textsc{e}arth, with the precision in  \unit[]{$\mu$m} \cite{tapley}. The precision of the estimated gravity field solution computed from the range-rate measurements has not met requirements defined by \cite{kim2000} before the launch of the \textsc{grace} mission. This requirement is termed as \textsc{grace} baseline which is still several orders of magnitude below the current achieved precision (cf. \textsc{f}ig. \ref{figBaseline}).

Out of many error sources responsible for this limited precision, one of the source is the  errors of the attitude sensors which propagate to the estimated gravity field solutions through the $\dot\rho$. In order to minimize them, we must know the characteristics of those attitude errors which are affecting the precision of the $\dot\rho$.
In the \textsc{grace} mission, so far, the model of the attitude errors is also unknown \cite{kim2000} which makes it even more difficult to analyze them as we do not know the frequency range where these errors dominates. One  way to analyze the attitude errors in this situation would be -- to compare the impact of different attitude datasets on the $\dot\rho$. Now, it is again difficult to analyze the direct impact of attitude errors on the $\dot\rho$ as they contain mass change signals and errors both, which makes it complicated to analyze the errors in details. Therefore, we analyze the attitude errors in the residuals of the $\dot\rho$ observations which we compute after the gravity field parameter estimation using least squares estimation as shown in Eqn. \ref{eq:1} \cite{koch,torstenphd}. These range-rate residuals ($\mathbf{\hat{e}}$ in Eqn. \ref{eq:1}) reflect the   errors which are partially absorbed by the estimated gravity field parameters ($\mathbf{\hat{x}}$). Thus, their analysis is a good basis to understand the attitude errors affecting the range-rate observations and the gravity field parameters. Therefore, in this contribution our aim is to present the results on --
\begin{enumerate}
  \item[\emph{A}.]  An analysis of the attitude error characteristics by comparing the two different attitude datasets
  \item[\emph{B}.]  The propagation of these errors into the K-band range-rate observations by analyzing their residuals.
\end{enumerate}

\subsection*{Details of the attitude data used in this work}

Before discussing the results of our findings, we discuss the datasets representing the \textsc{grace} satellite attitude used in this work --
\begin{itemize}
  \item[\textbf{\#1}] -- \textsc{sca1b} release 2, the standard Level1B attitude data computed from the combination of the data of the two star cameras present on each of the two spacecrafts \cite{l1bHbook}. The data of the two star cameras is combined using the algorithm described in \cite{romans}.
  \item[\textbf{\#2}] -- Reprocessed attitude (\textsc{sca1b} quaternions + \textsc{acc1b} angular accelerations), the set of quaternions provided in the \textsc{sca1b} are combined with the angular accelerations provided in the \textsc{acc1b} product. The combination details are provided in \cite{fusion}.
  For details about the \textsc{l}evel1\textsc{b} products (\textsc{sca1b} and \textsc{acc1b}), refer to  \cite{l1bHbook}.
  Here we present the results of the analysis of two years of \textsc{grace} data i.e. 2007 and 2008.
\end{itemize}

The attitude data is required to compute the range-rate antenna offset corrections (\textsc{aoc}) which are added to the \textsc{k}-band range-rates from the instrument frame to the \textsc{l}ine of \textsc{s}ight (\textsc{los}) frame of reference (cf. Eqn. \ref{eq1}). The \textsc{aoc} are computed as given in \cite{bandi15}.
\begin{align}
\dot{\rho}_{\text{LOS}} &= \dot{\rho}_{\text{KBR}} + \dot{\rho}_{\text{AOC}}
\label{eq1}
\end{align}
The attitude information is also needed to rotate the linear accelerations in \textsc{acc1b} product from the science reference frame to the inertial frame of reference \cite{l1bHbook}.
Thus, the attitude errors propagate to the range-rate observations via \textsc{aoc} and the linear accelerations which are used to reduce the effect of the drag acting on the spacecraft and then on the \textsc{k}-band observations. In this contribution, we discuss the propagated errors via \textsc{aoc}.
The range-rate residuals ($\mathbf{\hat{l}}$) obtained after removing all the perturbations from them are used as the observations in the gravity field parameter estimation (cf. Eqn. \ref{eq:1}). Once we estimate the gravity field parameters, the residuals ($\mathbf{\hat{e}}$) obtained are used in this study to analyze the attitude errors.
%
\begin{align}
  \mathbf{\hat{l}} - \mathbf{A} \mathbf{\hat{x}} = \mathbf{\hat{e}}
  \label{eq:1}
\end{align}
where, $\mathbf{\hat{l}}$ are the estimated range-rate observations,  $\mathbf{A}$ is the design matrix, $\mathbf{\hat{x}}$ contains the estimated gravity field parameters and  $\mathbf{\hat{e}}$ are the range-rate residuals estimated after least-squares fit using the \textsc{itsg-}2014 gravity field processing chain \cite{itsg2014}. 




\section{Results}
\subsection{Error characteristics of the two attitude datasets}
\begin{figure}[htbp!]
  \centering
\includegraphics[width = 1\linewidth]{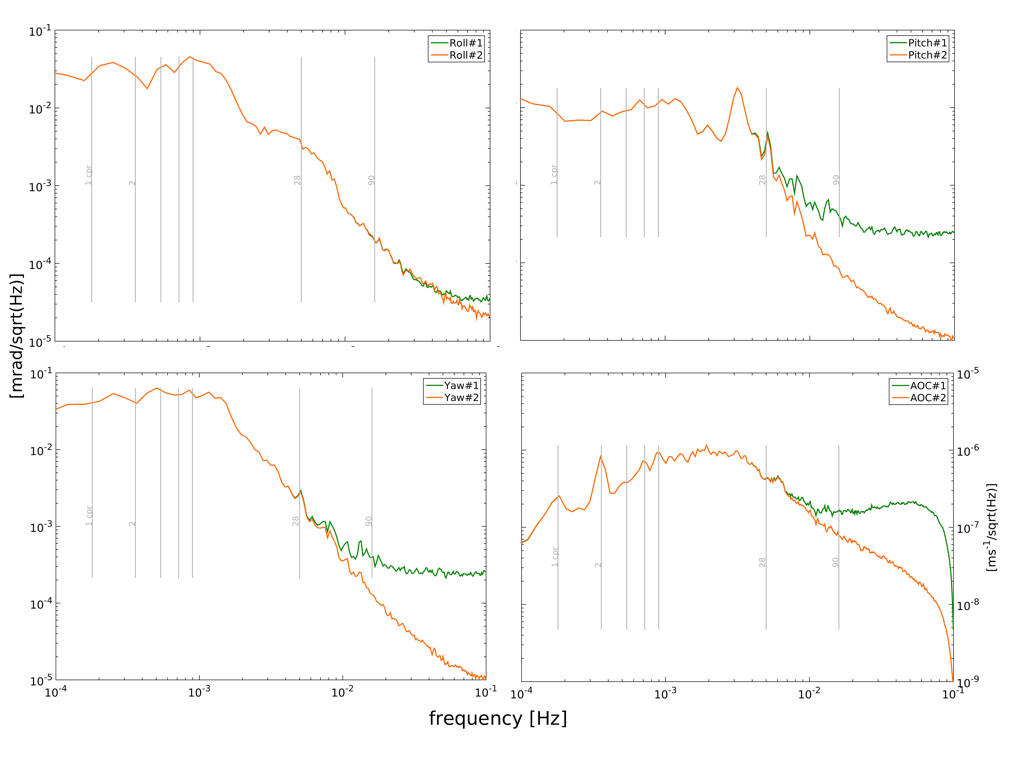}
\caption{PSD of the pointing angles computed from the two attitude datasets on the day 1, \textsc{d}ecember 2008. \label{fig1}}
\end{figure}

We compare the characteristics of the two attitude datasets by analyzing the differences between their pointing angles.  In \textsc{f}ig. \ref{fig1}, the comparison between the power spectral densities of the pointing angles roll, pitch and yaw of the attitude data \textbf{\#1} and \textbf{\#2} shows very small differences in roll, that too in very high frequencies whereas, the pitch and yaw angles shows the deviation after $>$\unit[5.5]{m\textsc{h}z} which becomes very large after $>$\unit[9]{m\textsc{h}z}. The pitch and yaw angles computed from the attitude data \textbf{\#2} have low high frequency noise which is due to the angular accelerations combined with the star camera datasets. Thus, the low accuracy of the star camera data in high frequencies, can be complimented by combining it with the angular accelerations.
Thus, we have shown that the combination of the angular accelerations with the star camera data improves the accuracy of the attitude in high frequencies.

Further we are interested in understanding the details of the characteristics of the yaw and pitch angles of attitude which are improved when star camera data is combined with the angular accelerations.
In order to analyze these characteristics, we plot the observations representing the pointing angles along the \emph{argument of latitude} and the time in days for the two years (i.e. 2007 and 2008). The values start from $0^{\circ}$ (bottom) to $360^{\circ}$ (top) for one complete orbit. The ticks on the y-axis represents -- the north (\textsc{np}), south poles (\textsc{sp}) and, the ascending \textsc{ae} and descending equator (\textsc{de}), respectively. The x-axis respresents the two years of time in terms of days. Further details about the \emph{argument of latitude} can be found in \cite{gill}.

%
%
%

\begin{figure}[htbp!]
  \centering
\includegraphics[width = 1\linewidth]{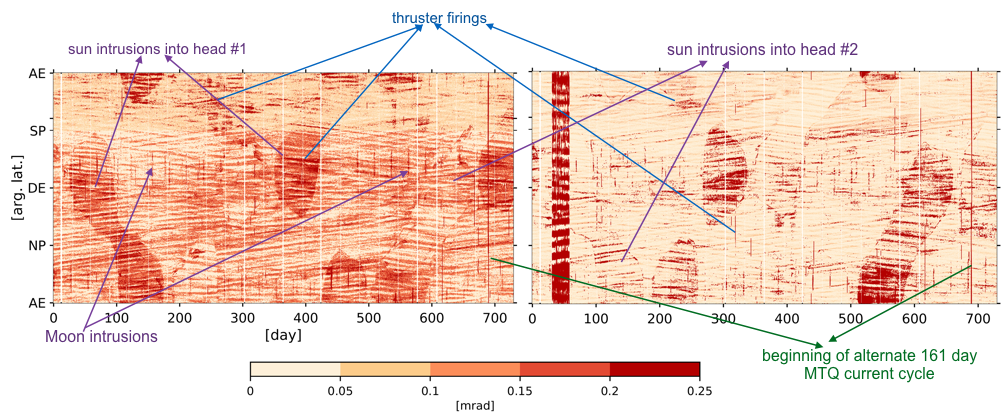}
\includegraphics[width = 1\linewidth]{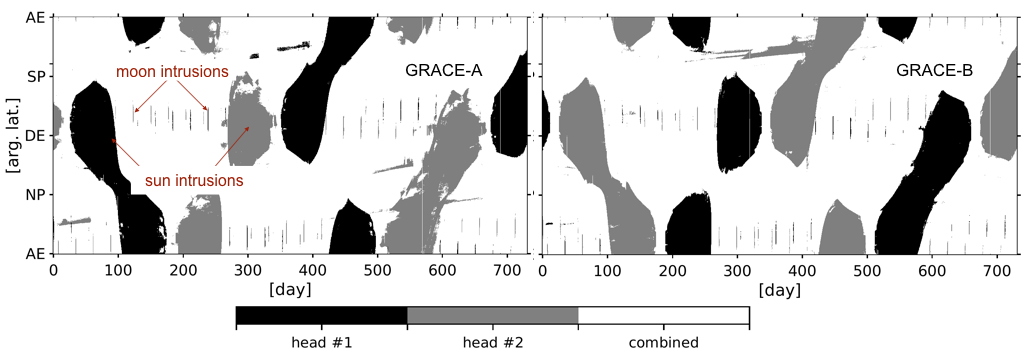}
\caption{\emph{top panel}: Absolute differences between the  pitch angles of \textsc{grace-a} and \textsc{grace-b}, computed from \textbf{\#1} and \textbf{\#2} respectively. \emph{bottom panel}: The data flags showing the blinding status of each star camera head of  \textsc{grace-a} and \textsc{grace-b}. \emph{left panel}: \textsc{grace-a} and \emph{right panel}: \textsc{grace-b}.
\label{fig3}}
\end{figure}

The differences between the pitch angles computed from the data \textbf{\#1} and \textbf{\#2} are
plotted in \textsc{f}ig. \ref{fig3}\emph{top panel},  shows that the attitude data \textbf{\#2} specially complement during the time when the star camera attitude is computed from only one available star camera head as shown in \textsc{f}ig. \ref{fig3}\emph{bottom panel}. The attitude data \textbf{\#1} is computed by combining the data of two star camera heads present on each satellite \cite{romans} when the data from two star cameras is available, otherwise, the attitude used is from one of the two available star camera heads. The availability of the star camera heads on each spacecraft can be seen in \textsc{f}ig. \ref{fig3}\emph{bottom panel}. The star cameras onboard gets blinded periodically by the \textsc{s}un (every \unit[161]{d}) and \textsc{m}oon (every \unit[21]{d}) intrusions into their field of view, thus, almost \unit[50]{\%} of the time, one of the two star cameras is blinded and attitude is obtained from another available star camera.
The attitude from only one star camera has high anisotropic errors as compared to the combined attitude solution \cite{bandi15}.
%
%

The differences in the pitch angles of two \textsc{grace} satellites are exceptionally high where, the attitude \textbf{\#1} is computed from only one star camera head. Thus, its combination with the angular accelerations reduces the errors significantly by providing more redundant information of the spacecraft's attitude. Therefore, the signatures related to the sun and moon intrusions into star cameras' field of view are clearly visible in the differences between the two pitch angles. Similarly, the differences between the yaw angles show the similar improved features (not shown here).  These differences indicate the improvements in the high frequencies shown in the \textsc{psd}s of the pointing angles shown in \textsc{f}ig. \ref{fig1}.

Besides this, the time period where the attitude \textbf{\#1} was affected by the attitude control actuators such as -- thruster firings and changes in the currents flowing through the magnetic torquer rods (\textsc{mtq}s), the differences are again high in \textsc{f}ig. \ref{fig3}\emph{above}. Some of the high difference places which are due to attitude actuators are marked in the \textsc{f}ig. \ref{fig3}\emph{above}.

Also, we observe that the high differences shown in the pitch angles  are not always consistent with the sun and moon intrusions into the star camera field of view. For example, in \textsc{grace-a}, when \emph{head\#1} was blinded from days 20 to 180, the differences in the pitch angles  were relatively high than the days from 195 to 250, where \emph{head\#2} was blinded. Based on the investigation of the accuracies of the two star camera heads onboard each spacecraft by \cite{herman,bandi15}, we know that the attitude delivered from the \emph{head\#2}  was more accurate than the \emph{head\#1} on both of spacecrafts.
Also, we know that there are periods where the attitude product \textsc{sca1b} is computed by the data of single available star camera head. We observed that when the \textsc{sca1b} was computed from the \emph{head\#2} only and this product is combined with the angular accelerations to produce attitude data \textbf{\#2}, the resulted data \#2 is more accurate than the attitude product \textbf{\#2} which is computed with the combination of attitude data from \emph{head\#1} only.

Thus, we see high differences in the pitch angles when attitude product is combined with the attitude based on the \textsc{sca} \emph{head\#2} and small differences when the attitude product is combined with the \textsc{sca} \emph{head\#1} data.
Therefore, we can say that the combination of the more accurate star camera data with the data of other sensors' (for example, angular accelerations in our case) leads to a more improved product. The accuracy of each attitude data set is also one of the important factors in the combination.

The places where the attitude \textsc{sca1b} product was computed by the available attitude from the two star camera heads, its differences from the reprocessed product are small which shows that the attitude based on star camera data only is also comparably accurate provided that the data of both star camera heads is available.


%
%

\subsection{Propagation of attitude errors to the K-band range-rate observations}
\begin{figure}[htbp!]
  \centering
\includegraphics[width = 0.79\linewidth]{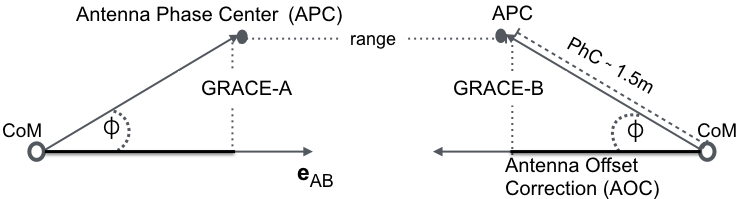}
\caption{A diagrammatic representation of the \textsc{aoc} applied to the \textsc{grace} \textsc{k}-band range-rate observations. \label{figAOC}}
\end{figure}

The above investigated errors in the attitude datasets propagate to the range-rate observations via \textsc{aoc} as shown in the Eqn. \ref{eqAOC}.
\begin{align}
\text{AOC} &= \text{PhC } cos\phi = e_{\text{AB}} . (\mathbf{R}_{\text{SRF, A}}^{\text{IRF}} \text{PhC}_{\text{A}}) - e_{\text{AB}} . (\mathbf{R}_{\text{SRF, B}}^{\text{IRF}} \text{PhC}_{\text{B}})
\label{eqAOC}
\end{align}
where, $\text{PhC}_{\text{A}}$  and $\text{PhC}_{\text{B}}$ are the distance from the \textsc{k}-band phase center to the satellite's center of mass, $e_{\text{AB}}$ is the position vector pointing from the spacecraft \textsc{a} to \textsc{b} (cf. \textsc{f}ig. \ref{figAOC}) and the rotation matrix  $\mathbf{R}_{\text{SRF}}^{\text{IRF}}$ representing the rotation from the \textsc{s}cience \textsc{r}eference \textsc{f}rame (\textsc{srf}) to the \textsc{i}nertial \textsc{r}eference \textsc{f}rame (\textsc{irf}), is computed from the quaternions representing attitude.

In \textsc{f}ig. \ref{fig1}, when we compare the \textsc{psd}s of the pointing angles with the \textsc{aoc}, the differences in the two sets of \textsc{aoc} are visible after frequency \unit[5.5]{m\textsc{h}z}. The two \textsc{psd}s of the \textsc{aoc} show large deviation in the high frequencies similar to the pitch and yaw angles.

Further, to investigate the differences between two sets of \textsc{aoc} computed from the attitude data \textbf{\#1} and \textbf{\#2}, we plot the observations on the \emph{argument of latitude} and time plots as shown in \textsc{f}ig. \ref{fig3}.
The differences between the two sets of \textsc{aoc} are correlated with the  differences between the pointing angles of the two \textsc{grace} spacecrafts as we can see in \textsc{f}ig. \ref{fig3}\emph{top panel}. In the \textsc{aoc} differences, we can see that the differences are high at the places when sun and moon intrudes into the star camera field of view. It indicates that the accuracy of the attitude data is highly limited by the intrusions blinding the star cameras field of view. Again, the high differences can be clearly seen and are consistent with the pitch angle differences, where the attitude is affected by the actuators actuated to control the spacecraft's attitude.

%
\begin{figure}[htbp!]
  \centering
  \includegraphics[width = 0.75\linewidth]{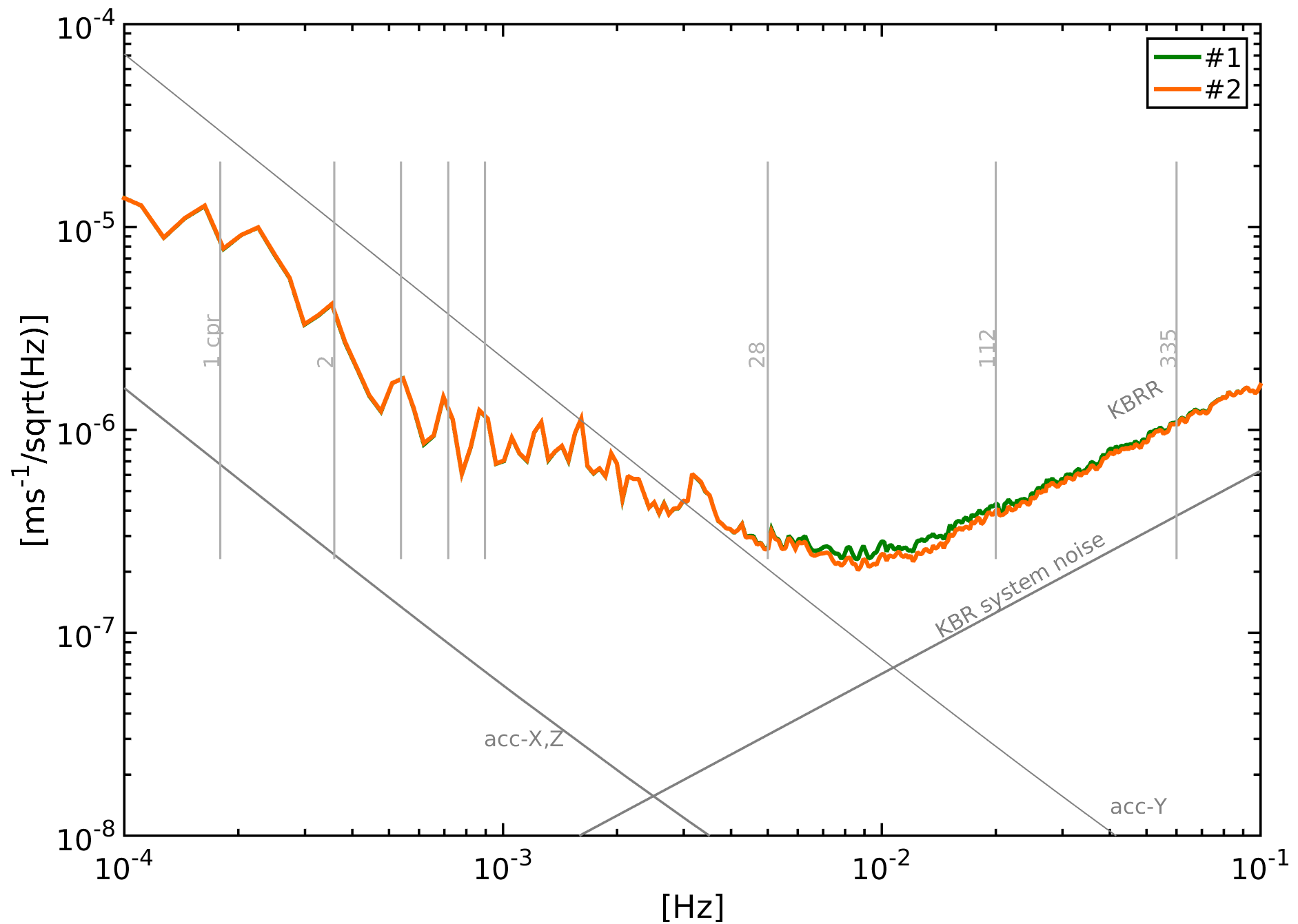}
  \caption{A comparison of the \textsc{psd} of the range-rate residuals computed from the two attitude datasets respectively.}
  \label{figPSDkbrr}
\end{figure}
\begin{figure}[htbp!]
  \begin{tabular}{m{0.0001\linewidth} m{0.999\linewidth} m{0.0001\linewidth}}
        \hspace{0.4cm}\vskip-0.9cm\rotatebox{90}{$\hat{e}_{\#2} - \hat{e}_{\#1}$} &
        \includegraphics[width = 1\linewidth]{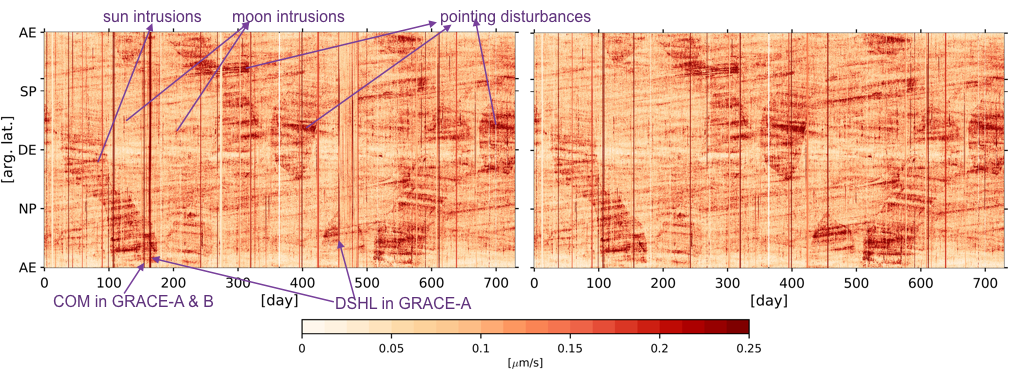} &
        \vskip-0.6cm\hspace{-0.45cm}\rotatebox{90}{$\text{AOC}_{\#2}-\text{AOC}_{\#1}$}
  \end{tabular}
\caption{\emph{left}: Absolute differences between the range-rate residuals; \emph{right}: Absolute differences between the \textsc{aoc} computed from \textbf{\#1} and \textbf{\#2}. \label{fig4}}
\end{figure}
There are high amplitude of residuals continuous over a full orbit, seen as vertical stripes which are mainly due to the satellite orbit and attitude control maneuvers (for example \textsc{c}enter of \textsc{m}ass calibration (\textsc{c}o\textsc{m}), yaw axis turn, thruster firings, large magnetic torquer rod currents) and heating table related changes (so called \textsc{dshl} events \cite{beerer}) which indirectly affect the attitude sensors, hence, their observations. The combined attitude data \textbf{\#2} improves the attitude which is affected due to such  maneuvers disturbances.
\begin{figure}[htbp!]
  \hspace{-0.4cm}
  \begin{tabular}{m{0.5\linewidth} m{0.5\linewidth}}
\includegraphics[width = 0.95\linewidth]{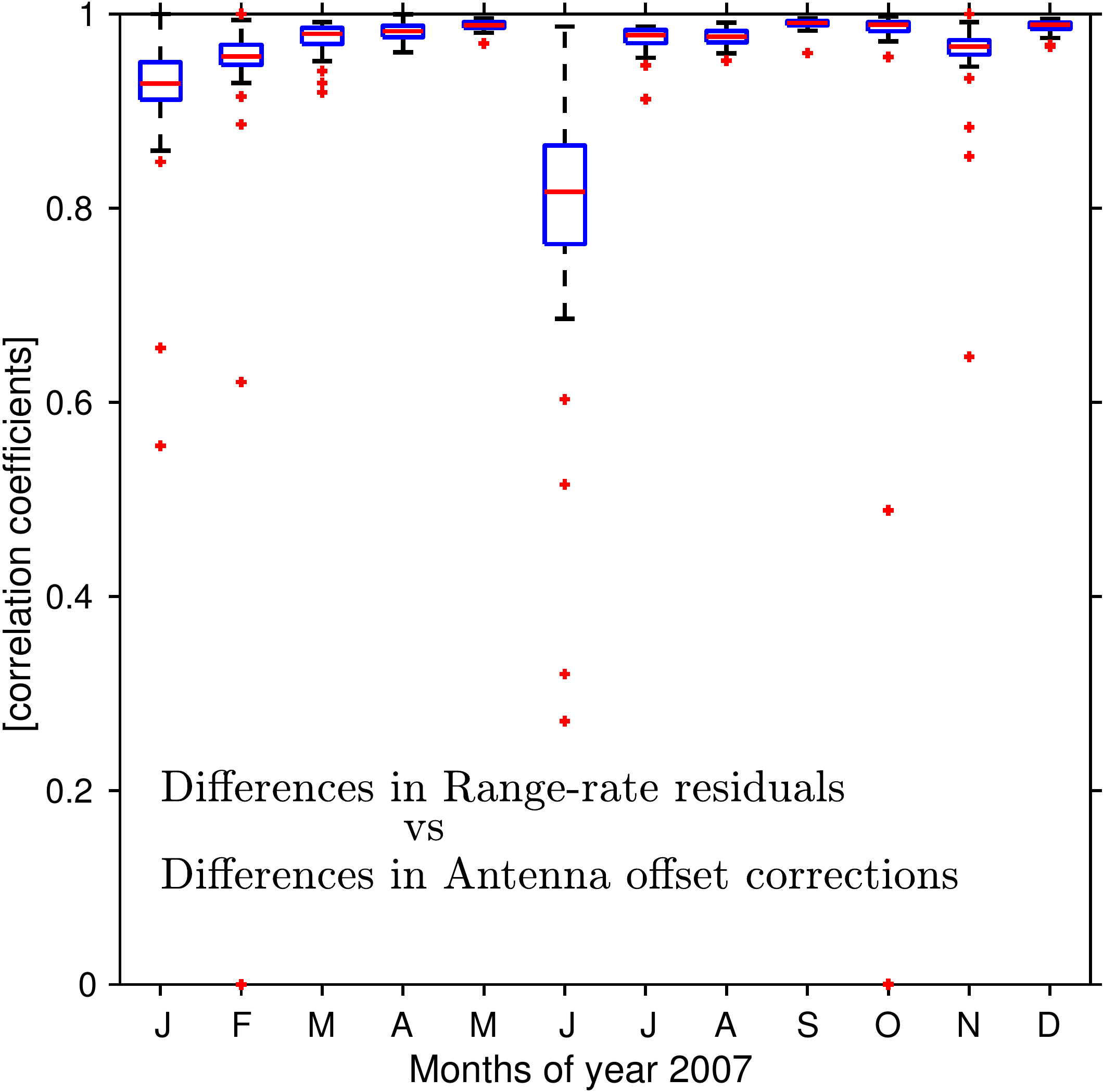} &
\includegraphics[width = 0.95\linewidth]{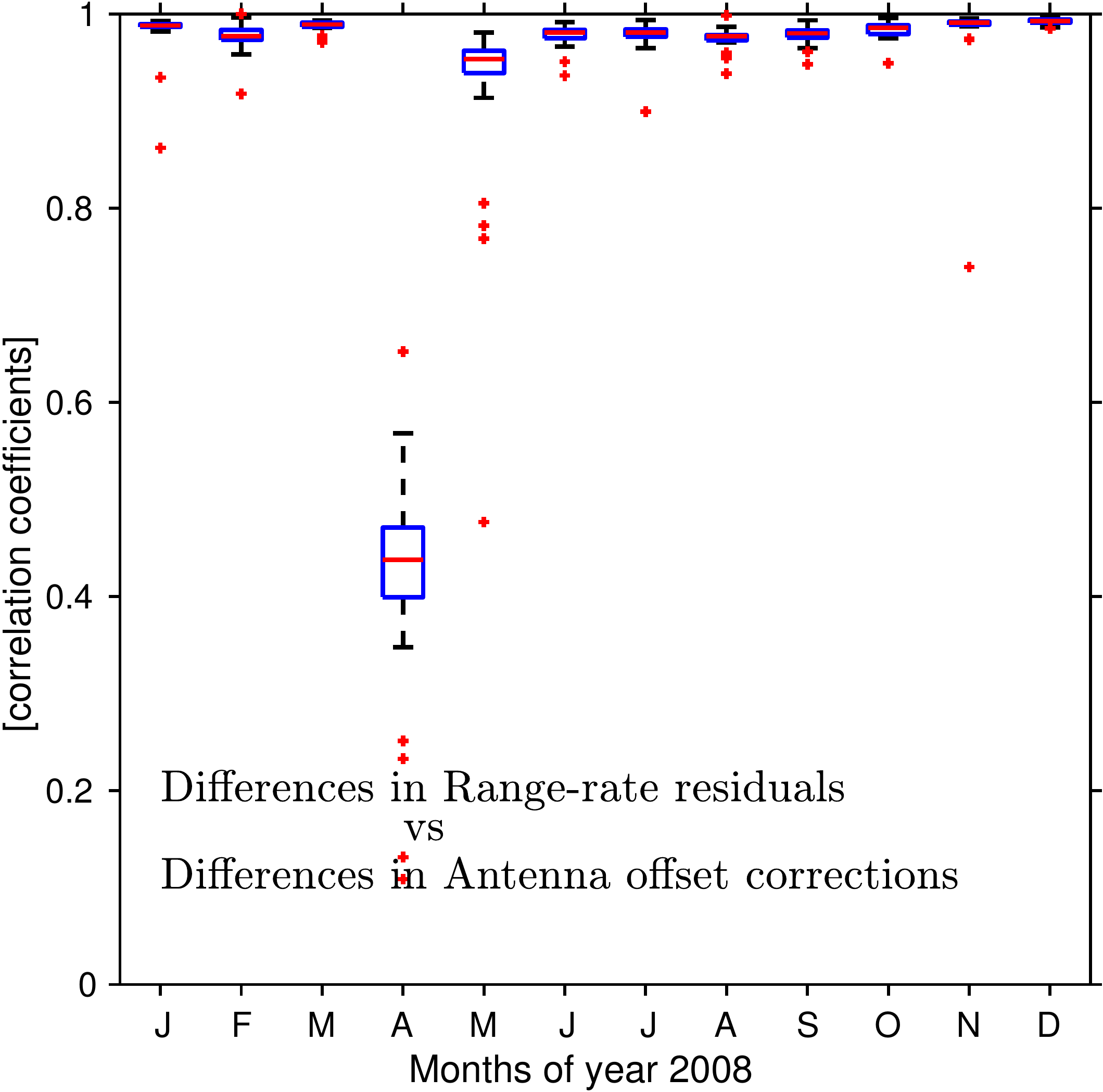}
  \end{tabular}
\caption{Correlations between the range-rate residual differences and the \textsc{aoc} differences (cf. \textsc{f}ig. \ref{fig4}) in monthly box plots for the two years. \label{fig5}}
\end{figure}

The \textsc{aoc} when added to the range-rate observations, propagate the attitude errors to the \textsc{k}-band range-rate observations (cf. Eqn. \ref{eq1}).  The presence of these errors in the range-rate observations may affect the quality of the gravity field solutions which is the end product estimated using the range-rate observations.

An analysis of the range-rate residuals should reveal these errors, thus, indicates an insufficiency in the  approach of handling the observation noise in the gravity field parameter estimation (cf. Eqn. \ref{eq:1}).
Therefore, we analyze the range-rate residuals computed after the least squares fit using each of the attitude dataset respectively. We represent the range-rate residuals as $(\hat{e}_{\#1})$ and $(\hat{e}_{\#2})$ computed from the attitude datasets \textbf{\#1} and \textbf{\#2} respectively. Now, when we compare the PSDs of two sets of residuals as shown in \textsc{f}ig. \ref{figPSDkbrr}, we observed that the two \textsc{psd}s deviate from the frequency \unit[5.5]{m\textsc{h}z} similar to the pitch and yaw angles. This indicates that the pitch and yaw errors are propagated to the range-rate residuals which affects the frequencies starting from \unit[5.5]{m\textsc{h}z}. However, we do not see the large deviations in the  high frequency ($>$\unit[10]{m\textsc{h}z}) range-rate residuals. It is due to the noise from other known sources which is the so called \textsc{kbr} instrument system noise and is also the dominated noise in the residuals \cite{thomas}. Thus we analyze the differences between the two set of residuals. The analysis shows the differences between the residuals are perfectly correlated with the differences of \textsc{aoc} between the two attitude datasets. The perfect correlations between the two sets of differences can be seen in Figs. \ref{fig4} and  \ref{fig5} respectively. The differences when plotted along the \emph{argument of latitude} and time, shows that their values are high when the attitude is  affected by the sun into one of the star cameras field of view and the affected attitude largely due to the actuators' actuated to control the attitude of the satellite. When we compute the correlation coefficients for daily observations for each month, we find that almost in every month the correlations between the \textsc{aoc} differences and residuals differences are close to 1 which shows the perfect correlation except for the month of \textsc{a}pril 2008 where correlation coefficients are very small, it may be due to the differences of residuals are propagated via the accelerometers, the discussion about it is out of scope of this paper.
\section{Conclusion}
We have presented the first results of the reprocessed attitude data (\textbf{\#2}) analysis with respect to the standard \textsc{sca1b} attitude data, provided by \textsc{jpl}. We show that the reprocessed attitude computed by combining the angular accelerations along with the star camera datasets certainly improves the overall attitude quality. It especially compliments during the time period where the standard attiutde has been computed from one star camera data only. Also, it reduces the errors in the standard attitude where the standard attitude is affected by the attitude actuators. However, the accuracy of the star camera data is an important factor has to be considered while combining it with other attitude sensors. High accurate star cameras lead to more accurate reprocessed combined attitude which has significantly less high frequency noise as compared to the combined data computed with less accurate star camera data.
Thus, we expect that the suggested  improvement of the star camera data by \cite{harvey} and its combination with the angular accelerations will further improve the remaining errors in the \textsc{grace} attitude data.

The attitude data \textbf{\#2} significantly reduces the pitch and yaw errors and correspondingly improves the \textsc{aoc}. We also noticed that the \textsc{aoc} is largely affected by the pitch and yaw pointing errors of the spaceacraft's attitude. Thus, they propagate to the range-rate observations via the \textsc{aoc} as shown in Eqn. \ref{eq1}.

The pitch and yaw errors largely propagates to the residuals which are revealed by their differences. The similar magnitude of the differences between the \textsc{aoc} and the range-rate residuals shows that the attitude errors largely propagate via \textsc{aoc} which is also proved by their correlation coefficients.

\section*{Acknowledgment}
\addcontentsline{toc}{section}{Acknowledgment}
\scriptsize
We acknowledge support from the \textsc{g}erman \textsc{r}esearch \textsc{f}oundation \textsc{dfg}
within \textsc{sfb} 1128 geo-\textsc{q} to fund this research. We would like to thank Prof. \textsc{j}akob \textsc{f}lury for the fruitful discussions with him in regard to this work.

\normalsize

%
%

%
%

\end{document}